\documentclass[twocolumn,showpacs,preprintnumbers,amsmath,amssymb]{revtex4}

\usepackage{graphicx}
\usepackage{dcolumn}
\usepackage{bm}

\def\ii{{\mathrm{i}}}

\def\dd{{\mathrm{d}}}

\def\sub#1{_{\mathrm{#1}}}
\def\up#1{^{\mathrm{#1}}}
\def\Vec#1{\mathbf{#1}}

\def\sumint{\hbox{$\sum$}\!\!\!\!\!\!\int}
\def\He4{{$^4$He}}

\begin{document}

\preprint{APS/123-QED}

\title{Localization of Bose-Einstein Condensation and Disappearance of Superfluidity of Strongly Correlated Bose Fluid in a Confined Potential}

\author{Michikazu Kobayashi}
\author{Makoto Tsubota}%
\affiliation{Faculty of Science, Osaka City University, Sugimoto3-3-138, Sumiyoshi-ku, Osaka558-8585, Japan}%


\date{\today}

\begin{abstract}
We develop a Bose fluid model in a confined potential to consider the new quantum phase due to the localization of Bose-Einstein condensation and disappearance of superfluidity which is recently observed in liquid \He4 in porous glass at high pressures. A critical pressure of the transition to this phase can be defined by our new analytical criterion of supposing the size of localized Bose-Einstein condensate becomes comparable to the scale of confinement. The critical pressure is quantitatively consistent with observations without free parameters.
\end{abstract}

\maketitle

\section{Introduction} \label{sec-intro}

A Bose condensed system in a random environment is a very important problem not only on the interests in the impurity effect but also for clarifying the long-range correlation, for example, the relation between Bose-Einstein condensation (BEC) and superfluidity. Experimental works have studied this system by means of liquid \He4 in porous glass and observed many interesting phenomena \cite{Reppy,Plantevin,Yamamoto}. In particular, Yamamoto {\it et al.} recently observed the disappearance of superfluidity of liquid \He4 confined in porous Geltech silica with the very small pore size 25\AA\ at high pressures and very low temperatures, suggesting the new quantum phase transition to localized BEC under the effect of strong correlation \cite{Yamamoto}.

In this work, motivated by their experiment, we investigate the behavior of a Bose fluid under the effect of confinement at zero temperature. We use the three-dimensional Bose fluid model in a confined potential \cite{Kobayashi}, and introduce our new criterion for finding the localization of BEC. Supposing a BEC is localized, we calculate the energy of the system as a function of the size of the localized BEC, and minimize the energy. The resulting size becomes the order of the scale of confinement above a critical pressure, when the superfluid density disappears. The critical pressure is quantitatively consistent with the experimental one without free parameters.

\section{The model of Bose fluid in a confined potential} \label{sec-model}

The Grand canonical Hamiltonian $\hat{H}-\mu\hat{N}$ of a Bose fluid in a confined potential is given by
\begin{widetext}
\begin{equation}
\hat{H}-\mu\hat{N}=\sum_{\Vec{k}}[\varepsilon(\Vec{k})-\mu]\hat{a}^\dagger(\Vec{k})\hat{a}(\Vec{k})+\frac{1}{V}\sum_{\Vec{k}_1, \Vec{k}_2}U(\Vec{k}_1-\Vec{k}_2)\hat{a}^\dagger(\Vec{k}_1)\hat{a}(\Vec{k}_2)+\frac{1}{2V}\sum_{\Vec{k}_1, \Vec{k}_2, \Vec{q}}g_0(\Vec{q})\hat{a}^\dagger(\Vec{k}_1+\Vec{q})\hat{a}^\dagger(\Vec{k}_2-\Vec{q})\hat{a}(\Vec{k}_2)\hat{a}(\Vec{k}_1), \label{eq-hamiltonian}
\end{equation}
\end{widetext}
where $\hat{a}^\dagger(\Vec{k})$ and $\hat{a}(\Vec{k})$ are the free Boson creation and annihilation operators with the wave number $\Vec{k}$, $\varepsilon(\Vec{k})=\hbar^2\Vec{k}^2/2m$ is the kinetic energy of a particle of mass $m$, $\mu$ the chemical potential. $U(\Vec{k})$ is the external confined potential, $V=L^3$ the volume of the system having the size $L$ and $g_0(\Vec{k})$ the interaction between two particles. The second term of the right-hand side represents the interaction between one particle and the confined potential; the second-order perturbation yields the Green function
\begin{equation}
  G\up{R}(k)=\frac{1}{\hbar^2V^2}\sum_{\Vec{k}_1}|U(\Vec{k}-\Vec{k}_1)|^2G^0(\omega,\Vec{k})^2G^0(\omega,\Vec{k}_1).\label{eq-potential-perturbation}
\end{equation}
The third term refers to the interparticle interaction. Using the ring-approximation, we obtain
\begin{equation}
  g(q)=\frac{g_0(\Vec{q})}{\displaystyle 1+\frac{g_0(\Vec{q})}{\hbar V}\sumint_{\Vec{k}}\frac{\dd k_0}{2\pi\ii}\:G^0(q)G^0(k+q)}.\label{eq-interaction-perturbation}
\end{equation}
Here $G^0(k)$ is the noninteracting Green function
\begin{equation}
  G^0(k)=G^0(\omega,\Vec{k})=\frac{\hbar}{\hbar\omega-[\varepsilon(\Vec{k})-\mu]},\label{eq-Green-function-0}
\end{equation}
with the frequency $\omega$ and $k=(\omega,\Vec{k})$. Following the Bogoliubov method, we separate $\hat{a}(\Vec{k})$ to the condensed part $a(\Vec{k}\sub{c})=\sqrt{N\sub{c}}$ with the smallest wave number $\Vec{k}\sub{c}=(2\pi/L,2\pi/L,2\pi/L)$, and the noncondensed part $\hat{a}(\Vec{k}\neq\Vec{k}\sub{c})$. Then we calculate the Green function $G\up{I}(k)$ including interparticle interaction $g(k)$ by the Bogoliubov theory, finally obtaining the total Green function $G(k)=G\up{R}(k)+G\up{I}(k)$. Superfluid component $N\sub{s}$ based on the two fluid model can be given by the linear response theory \cite{Hohenberg}
\begin{equation}
  N\sub{s}=N-\frac{\hbar}{\displaystyle 6\pi m \ii}\sumint_{\Vec{k}}\dd k_0\:k^2\det[G(k)],\label{eq-superfluid-density}
\end{equation}
where $N$ is the total number of the particles. We assume the gaussian confinement $U(\Vec{k})=U_0\exp[-k^2/2k\sub{p}^2]$ so that the wave number $k\sub{p}=2\pi/r\sub{p}$ can be connected with the pore size $r\sub{p}$ of porous glass. Here the strength $U_0$ is estimated from the experimental critical coverage below which superfluidity vanishes \cite{Kobayashi}. For the interparticle interaction, we use $g_0(\Vec{k})=v_0(\sigma\sqrt{2\pi})^3\exp[-k^2\sigma^2/2]$, where parameters $v_0$ and $\sigma$ are determined by the comparison with the potential proposed by Aziz {\it et al.} \cite{Aziz}.

To consider the phase of localized BEC, we introduce a new criterion. First we assume a localized BEC of the size $L\sub{g}$ and calculate the energy $E\sub{g}$ by replacing the volume $V$ in Eq. (\ref{eq-hamiltonian}) with $V\sub{g}=L\sub{g}^3$. The total energy of the volume $V$ should be proportional to $E\sub{g}V/V\sub{g}$, because the number of the localized BECs is proportional to $V/V\sub{g}$. The ideal or weakly interacting Bose gas has $E\sub{g}$ proportional to $V\sub{g}$. In the strongly correlated Bose fluid, however, $E\sub{g}$ is no longer a simple linear function of $V\sub{g}$, then $E\sub{g}/V\sub{g}$ depends on $V\sub{g}$. Calculating $E\sub{g}/V\sub{g}$ as a function of $V\sub{g}$, we find $V\sub{g}=V\sub{g,min}$ which minimizes $E\sub{g}/V\sub{g}$. The localization of BEC can be defined by the new criterion. If $V\sub{g,min}$ exceeds $V$, the system is a non-localized BEC state. When $V\sub{g,min}$ is reduced to become comparable to $r\sub{p}^3$, we judge that the BEC is localized. Since $V\sub{g,min}$ is a function of $N$ and the pressure $P$, we can obtain the phase boundary of the transition to localization of BEC by this criterion.

\begin{figure}[btp]
\includegraphics[width=0.75\linewidth]{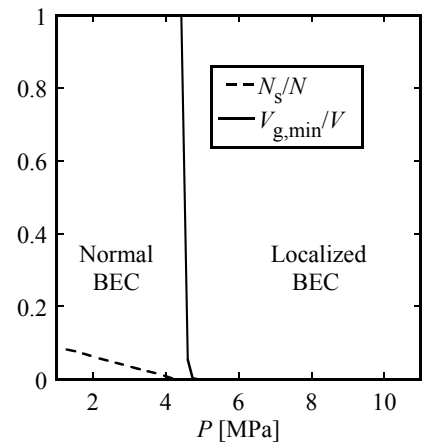}
\caption{\label{fig-localBEC} Dependence of the superfluid component $N\sub{s}$ and the volume $V\sub{g,min}$ of the localized BEC on the pressure. The pressure $P$ can be obtained by the thermodynamic relation $P=-\partial E/\partial V$.}
\end{figure}

Figure \ref{fig-localBEC} shows the dependence of $N\sub{s}$ and $V\sub{g,min}$ on the pressure, where all numerical parameters are fixed from Reference \cite{Yamamoto,Aziz} and remain no free parameters. At the pressure $P\simeq 4.2$ MPa, superfluid component $N\sub{s}$ disappears and $V\sub{g,min}$ is reduced dramatically to the order of $r\sub{p}^3$. We can, therefore, define this pressure as the critical pressure $P\sub{c}$.
Our critical pressure $P\sub{c}\simeq 4.2$ MPa is quantitatively consistent with the experimental one $P\sub{c}\simeq 3.5$ MPa \cite{Yamamoto} and we also conclude that experimental disappearance of superfluidity at high pressures is caused by the transition to localization of BEC. For the case of porous glass of the larger pore size $r\sub{p}\simeq 70$ \AA , we obtain the much larger critical pressure $P\sub{c}\simeq 9$ MPa which is too high for a BEC to be localized against solidification of liquid \He4, which is also consistent with experimental result for Vycor glass \cite{Yamamoto}.

\section*{Acknowledgments}
MT and MK both acknowledge support by Grant-in-Aid for Scientific Research (Grant No. 15340122 and 1505983 respectively) by the Japan Society for the Promotion of Science.

\end{document}